\documentclass[aps,prl,twocolumn,10pt,longbibliography]{revtex4-1}
\usepackage{amsmath}
\usepackage{amsfonts}
\usepackage{amssymb}
\usepackage{mathtools}
\usepackage[colorlinks=true,citecolor=blue,linkcolor=blue,urlcolor=blue]{hyperref}
\usepackage{bm}
\usepackage{times}
\usepackage{cases}
\usepackage{wasysym}
\usepackage[version=4]{mhchem}
\usepackage{graphicx}
\usepackage[caption=false]{subfig}
 
\begin{document}
\title{Quadrupole moments and their interactions in the triangular lattice antiferromagnet FeI$_2$}
\author{Gang Chen}
 \affiliation{International Center for Quantum Materials, School of Physics, Peking University, Beijing 100871, China}
\affiliation{Department of Physics and HKU-UCAS Joint Institute for Theoretical and Computational Physics at Hong Kong, 
The University of Hong Kong, Hong Kong, China}
     
\date{\today}
    
\begin{abstract}
Motivated by the recent experiments on the triangular lattice antiferromagnet FeI$_2$, we consider 
the presence of the quadrupole moments and their interactions between the spin-orbital-entangled 
local moments. In addition to the anisotropic pairwise interaction between the local moments, 
the interaction between the quadrupole moment arises from the interaction of the orbital occupation 
configurations. This is argued primarily from the time reversal symmetry and the different orbital 
configurations via their exchange paths. We discuss the implication of these interactions and expect 
this result to be complementary to the existing works in this system and other systems alike. 
\end{abstract}

\maketitle




%
Quantum many-body systems with a large local and active Hilbert space have 
attracted a significant attention in the recent years~\cite{chen2021mott}. 
These include the well-known Sachdev-Ye-Kitaev non-Fermi liquid model with a large number of 
interacting Majorana fermions~\cite{Chowdhury_2022}, 
the Kondo-lattice-like models for intermetallics with both local moments 
and itinerant electrons~\cite{Coleman}, the spin-orbital-coupled Mott insulators~\cite{Witczak_Krempa_2014}, 
even the weak Mott insulators with 
active charge and spin degrees of freedom~\cite{PhysRevLett.95.036403,PhysRevB.78.045109}, and so on. 
In these systems, there often exist a large number of degrees of freedom in the local Hilbert space. 
These degrees of freedom are sometimes of the same character, and more commonly, 
they are of quite distinct characters with very different physical properties. 
For the latter case, their differences may help people to clarify their self and mutual interactions
as well as the probing of their properties in the actual experiments. For example, 
in the intermetallic systems with the description of the Kondo-lattice-like models,  
the local moments could be well modeled as the spins and probed by usual magnetic 
measurements such as neutron scattering, while the itinerant 
electrons are more appropriately described by the Landau quasiparticles in the model setting 
and probed by the angle resolved photon emission, electric transports, et al. 
For the case of the spin-orbital-coupled Mott insulators that are of interest in this work, 
the distinct characters of the spin and the orbital can actually make many interesting differences
in the understanding of the candidate quantum materials.

The distinct characters between the spin and the orbital degrees of freedom have already been 
adopted in many early works in the field. 
The spin interaction without the involvement of the orbitals does not usually carry any anisotropy. 
With the intrinsic orientation dependence, the orbitals bring the spatial anisotropy into the model. 
One representative example of such models are the well-known Kugel-Khomskii spin-orbital exchange model~\cite{chen2021mott,Kugel_1982}. 
Upon the spin-orbital entanglement due to the strong spin-orbit coupling, the resulting effective spin models 
develops both the spatial anisotropy and the effective-spin-space anisotropy~\cite{PhysRevB.78.094403,Jackeli_2009,PhysRevB.82.174440}. 
These models 
have played an important role in Kitaev magnets~\cite{Jackeli_2009}, 
pyrochlore spin ice materials~\cite{PhysRevLett.105.047201,PhysRevLett.112.167203,PhysRevLett.98.157204}, 
iridates, osmates~\cite{PhysRevB.84.094420}, 
rare-earth triangular lattice magnets~\cite{PhysRevLett.115.167203,PhysRevB.94.035107,PhysRevB.96.054445}, 
and even many late $3d$ transition metal compounds
such as cobalts and vanadates~\cite{PhysRevB.97.014407,PhysRevB.97.014408,PhysRevLett.103.067205}. 
In this work, we are inspired by the recent work~\cite{Bai_2021} on the triangular lattice antiferromagnet FeI$_2$ 
and attempt to understand the interactions of spin and orbital degrees of freedom. 

\begin{figure}[b]
\includegraphics[width=8.5cm]{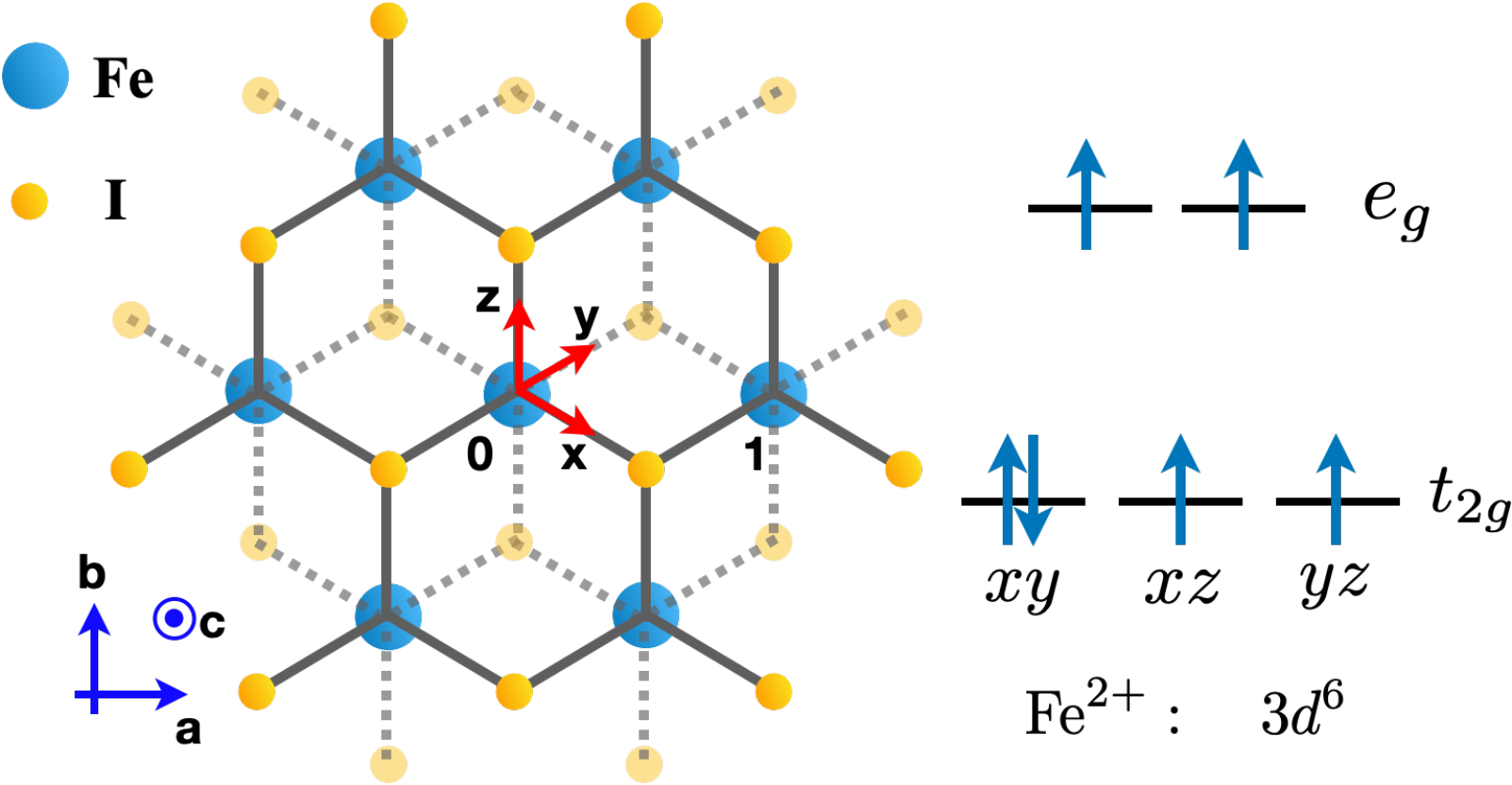}
\caption{(Color online.)
The left is the lattice structure with atomic positions of FeI$_2$. 
The bright (light) yellow spheres refer to the upper (lower) I atoms. 
The red axes ($x,y,z$) refer to the spatial coordinate system for the orbitals,
and the blue axes ($a,b,c$) refer to the coordinate system for the spin ${\boldsymbol S}$, 
the orbital angular moment ${\boldsymbol L}$,
and the effective spin moment ${\boldsymbol J}$. 
The right is the electron configuration for the Fe$^{2+}$ ion under the perfect 
octahedral environment in FeI$_2$. 
The orbital configuration in this plot is referred as the XY orbital occupation.
}
\label{fig1}
\end{figure}

In FeI$_2$, the Fe$^{2+}$ ion is located in an approximately octahedral environment.
Roughly, the orbitals are separated into the upper $e_g$ and the lower $t_{2g}$ orbitals,
and the electron occupation configuration is shown in Fig.~\ref{fig1}. 
Unlike the tetrahedral environment in the diamond lattice antiferromagnet FeSc$_2$S$_4$, 
the upper $e_g$ orbitals are half-filled without any orbital degeneracy.  
The lower $t_{2g}$ are filled with four electrons. 
In the case of the orbital degeneracy in the $t_{2g}$ manifold, the orbital degrees of freedom
are active, and the spin-orbit coupling plays a crucial role at the linear order. Even though the 
orbital degeneracy is actually lifted by the non-perfect-octahedral environment,
if the orbital separation is not large enough compared to the spin-orbit coupling, the spin-orbit coupling
is still able to entangle the spin and orbital degrees of freedom here.
From the Hund's coupling, the total spin is ${S=2}$. The triply degenerate $t_{2g}$ 
orbitals provide an effective orbital angular momentum ${L=1}$. 
The spin-orbit coupling entangles them together and leads to 
an effective spin moment ${\boldsymbol J}$ with ${J=1}$. 
In the modeling of Ref.~\onlinecite{Bai_2021}, the orbital character of the moment ${\boldsymbol J}$ 
has ready been considered, and the anisotropic effective spin interaction
was obtained by considering the symmetry operation on the local moments~\cite{Bai_2021,PhysRevB.94.035107}. 
Since the effective local moment is not a spin-1/2 moment, the 
local Hilbert space is a bit large, and the spin-orbital entanglement allows the 
system to more effectively access this large local Hilbert space~\cite{chen2021mott}.

Without the careful derivation of the actual model, one can make some progress 
by the microscopic and the symmetry analysis. We will take this approach and then
perform the calculation for the effective model. First of all, the effective model 
is obtained from a bit more parent model with the following form, 
\begin{eqnarray}
H = H_{\text{KK}} + H_{\text{SOC}} + H_{\text{ani}}, 
\label{eq1}
\end{eqnarray}
where the first term $H_{\text{KK}}$ refers to the Kugel-Khomskii spin-orbital exchange
with the ${S=2}$ and ${L=1}$ moment, the second term $H_{\text{SOC}}$ is the 
atomic spin-orbit coupling with the form of $\sum_i \lambda\, {{\boldsymbol L}_i \cdot {\boldsymbol S}_i}$,
and the third term $H_{\text{ani}}$ arises from the splitting among the $t_{2g}$ manifold. 
The second and third terms provide the local onsite Hamiltonian that determines the 
structure of the local moment. $H_{\text{SOC}}$ favors a ${J= |S-L|=1}$ local moment,
and $H_{\text{ani}}$ becomes the single-ion anisotropy for the $J$ representation. 
In the spirit of the degenerate perturbation theory, $H_{\text{KK}}$ is then projected
onto the degenerate manifold of the $J=1$ states on each lattice site, and 
reduced to the effective spin model. Since the Kugel-Khomskii spin-orbital exchange model is usually
a bit complicated, we here provide a physical understanding before performing the above procedure.

\begin{figure}[t]
\includegraphics[width=8.5cm]{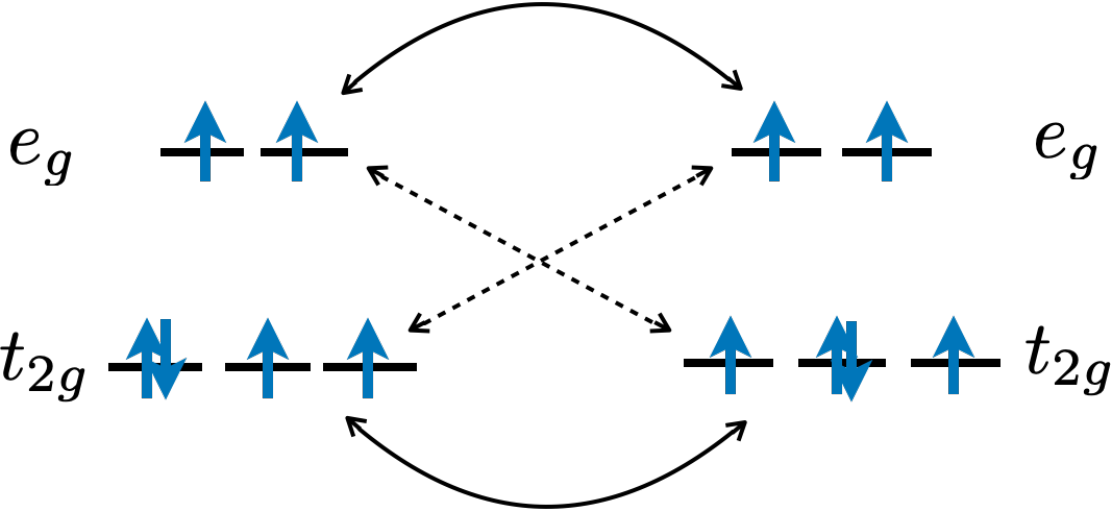}
\caption{(Color online.) The classification of 
the exchange process according to the manifold of the orbitals between two sites. 
The solid (dashed) arrows refer to the homo-orbital (hetero-orbital) exchange processes 
for the electrons from the identical (different) orbital manifolds. 
}
\label{fig2}
\end{figure}

First of all, there exist eight different Hermitian operators for the ${J=1}$ local moment. 
Among them, three are simple dipole moments, i.e. $J^{\mu}$ (${\mu =a,b,c}$),
and five quadrupole moments with
\begin{eqnarray}
&& Q^{3c^2} =  \frac{1}{\sqrt{3}} [ 3 (J^c)^2 - {\boldsymbol J}^2] , \\
&& Q^{a^2b^2}  =  (J^a)^2 - (J^b)^2 ,\\
&& Q^{ab} = \frac{1}{2} (J^a J^b + J^b J^a)  ,\\
&& Q^{ac} = \frac{1}{2} (J^a J^c + J^c J^a)  ,\\
&& Q^{bc} = \frac{1}{2} (J^b J^c + J^c J^b) . 
\end{eqnarray}
The single-ion anisotropy $(J^c_i)^2$ can be thought as an onsite polarization term for 
the quadrupole moment $Q^{3c^2}$. Because the quadrupole moments 
involve the product to two $J^{\mu}$ operators, they are even under time reversal symmetry. 
The dipole moments, however, 
are odd under time reversal symmetry. The distinction between the 
quadrupole and dipole moments may be cast as an example of  
magnetic moment fragmentation~\cite{PhysRevX.4.011007,PhysRevB.94.104430}.

The usual pairwise spin interaction, that is obtained by the symmetry analysis,
often includes the interaction between the dipole moments, and does not include
the quadrupole moments. To illustrate the origin of these complicated interactions,
we here return to the microscopic exchange model. 
 Unlike the above separation of the degrees of freedom into the spin part with $S=2$
 and the orbital part with $L=1$, we now separate the local moment from the upper $e_g$
 one and the lower $t_{2g}$ one. 
 The upper $e_g$ manifold provides a spin ${S_{e_g} =1}$ moment, while the lower 
 $t_{2g}$ manifold provides a spin ${S_{t_{2g}}=1}$ and an orbital angular moment ${L=1}$
 with an active spin-orbit coupling. 
 The $e_g$ spin and the $t_{2g}$ spin are then coupled with the ferromagnetic Hund's coupling,
 forming the total spin with ${S=2}$. In this picture, the local Hamiltonian is given as 
\begin{eqnarray}
\sum_i \big[{\bar{\lambda} {\boldsymbol S}_{i,t_{2g}} \cdot {\boldsymbol L}_i - J_H 
 {\boldsymbol S}_{i,t_{2g}} \cdot  {\boldsymbol S}_{i,e_g}} \big] + H_{\text{ani}} ,
\end{eqnarray} 
where 
 $J_H$ is the Hund's coupling,
and $\bar{\lambda}$ is related to the $\lambda$ in Eq.~\eqref{eq1} when 
$ {\boldsymbol S}_{i,t_{2g}} $ is reduced to the total spin ${\boldsymbol S}_i$ 
as ${\boldsymbol S}_{i,t_{2g}} \rightarrow {\boldsymbol S}_i /2$ after being 
symmetrized with ${\boldsymbol S}_{i,e_g}$. As it is equivalent to the previous discussion,
  this local Hamiltonian also gives the ${J=1}$ spin-orbit-entangled 
 local moment.

In the exchange of the electrons with both spin and orbital flavors for the Kugel-Khomskii model, 
one can classify the exchange process according to the orbital manifolds of the relevant electrons. 
This is depicted in Fig.~\ref{fig2}. One can make a couple interesting observation and simplify the 
Kugel-Khomskii exchange model, which we explain below.

For the homo-orbital exchange process for the electrons from the upper $e_g$ orbitals,
since the orbital sector is quenched for the $e_g$ orbital filling, this exchange process will simply give 
rise to the Heisenberg model in the form of
\begin{eqnarray}
\sim J_{1} {\boldsymbol S}_{i, e_g} \cdot {\boldsymbol S}_{j, e_g} .
\end{eqnarray}
This interaction, after projecting onto the ${J=1}$ local moment manifold, will lead to the 
pairwise interaction between the dipole moments. 

For the hetero-orbital exchange process between the electrons from the upper $e_g$ and lower $t_{2g}$ (see
the dashed line in Fig.~\ref{fig2}), the orbitals from the $t_{2g}$ manifold become active and contribute to the exchange
interaction. The site that contributes the $e_g$ electrons, however, does not carry an active orbital information. Thus,
the resulting exchange interaction should be of the following form,
\begin{eqnarray}
\sim J_2 \big[ {\boldsymbol S}_{i, e_g} \cdot {\mathcal O}_j +{\mathcal O}_i \cdot   {\boldsymbol S}_{j, e_g}    \big] ,
\label{eq9}
\end{eqnarray}
where ${\mathcal O}_j$ is a complicated operator that involves both ${\boldsymbol S}_{j,t_{2g}}$
and the orbital moment ${\boldsymbol L}_j$. Due to the time reversal symmetry,
the operator ${\mathcal O}_j$ should be odd under time reversal. Moreover,
even if the operator ${\mathcal O}_j$ may involve the operator like $i Q_{j}^{\mu}$ (with $Q_{j}^{\mu}$ as one of the quadrupole moments)
when it is projected to the ${J=1}$ manifold, the interaction between the dipole moment and the quadrupole moment should
 cancel out eventually by the hermiticity condition. 
 Thus, 
despite involving the complicated ${\mathcal O}_j$ operator, Eq.~\eqref{eq9} should also be reduced to the 
pairwise interaction between the dipole moments $J^{\mu}$ after projecting onto the $J=1$ local moment manifold.

\begin{figure}[t]
\includegraphics[width=7cm]{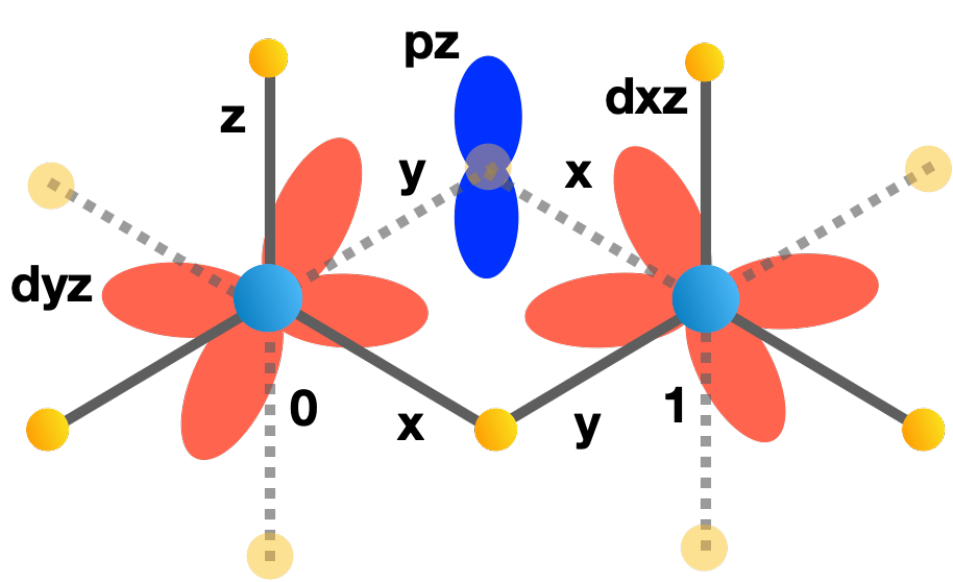}
\caption{(Color online.) The Fe-I-Fe superexchange path. 
In the plot, the relevant orbitals are the $d_{yz}$ orbital at Fe-0, 
the $p_z$ orbital at the intermediate I, and the 
$d_{xz}$ orbital at Fe-1. 
Likewise, along the equivalent path via the other I atom, the 
 the relevant orbitals are the $d_{xz}$ orbital at Fe-0, 
the $p_z$ orbital at the intermediate I, and the 
$d_{yz}$ orbital at Fe-1. 
}
\label{fig3}
\end{figure}

For the homo-orbital exchange process between the lower $t_{2g}$ manifold, both the spins and the orbitals
are involved. There are two types of operators that are involved in this exchange interaction. 
One type of operators is odd in $S_{t_{2g}}$ or $L$, such as $S_{i,t_{2g}}^{\mu}$
and $S_{i,t_{2g}}^{\mu} L_i^{\alpha} L_i^{\beta}$. 
These operators are labeled as $\mathcal{O}$. 
The exchange interaction between these $\mathcal{O}$ operators should be mostly reduced to the 
pairwise interaction between the dipole moments $J^{\mu}$ after projecting onto the ${J=1}$ 
local moment manifold. Since $\mathcal{O}$ could be related to the quadrupole moment as $i Q^{\mu}$,
there can still exists some part of the interaction that becomes the quadrupole-quadrupole interaction. This requires 
a careful calculation. 
The other type of operators is even in $L$, such as $L_i^{\alpha} L_i^{\beta}$. 
These operators are labeled as ${\mathcal N}$. Due to this even property under the time reversal,
the exchange interaction between these ${\mathcal N}$ operators should 
be mostly reduced to constant and/or the interaction between the quadrupole moments $Q$'s. 
Again, since ${\mathcal N}$ could be related to the dipole moment as $i J^{\mu}$, 
some part of the interaction could become the interaction between the dipole moments. 
Thus, to obtain the interaction between the quadrupole moments in this system, one essentially 
needs to keep track of the exchange interaction from the homo-orbital exchange process between 
the lower $t_{2g}$ manifold for both ${\mathcal O}$-${\mathcal O}$
and ${\mathcal N}$-${\mathcal N}$ interactions.

After the above classification and simplification, one can now focus on the 
exchange process between the $t_{2g}$ electrons. 
This exchange process will be identical to the one for the $d^4$ electron configurations, 
or equivalently, the $d^2$ hole configurations
on the $t_{2g}$ manifold. 
We single out the 
bond between the Fe site 0 and the Fe site 1 in Fig.~\ref{fig1}, and the exchange on
 the other bonds can be obtained by the symmetry operation. 
 With the assumption of the ideal geometry, the Fe-I-Fe exchange path is 90-degree. 
 Since here we are dealing with $3d$ electrons, we consider the indirect superexchange path
 via the intermediate I atoms. In principle, there could be a weak direct Fe-Fe exchange, for example,
via the $xy$ orbitals from the site 0 and the site 1 in Fig.~\ref{fig3}. 
For the indirect exchange path in Fig.~\ref{fig3}, there are two relevant hopping processes.
One is the $d_{yz}$-$p_z$-$d_{xz}$ from Fe-0 to Fe-1 via the upper I atom as we have depicted 
in Fig.~\ref{fig3}. The other is the $d_{xz}$-$p_z$-$d_{yz}$ from Fe-0 to Fe-1 via the lower I atom. 

Since the exchange of the $t_{2g}$ manifolds is equivalent to the one for the $d^4$ Mott insulators,
one can actually use the existing results for the superexchange interaction~\cite{PhysRevLett.111.197201}. 
Following Ref.~\onlinecite{PhysRevLett.111.197201}, we obtain the
 exchange interaction
for the Fe-0 and the Fe-1 that is then given as
\begin{eqnarray}
H_{t_{2g},0\leftrightarrow 1} &=& J_{\text{KK}} \Big[ 
\big( {\boldsymbol S}_{t_{2g},0} \cdot {\boldsymbol S}_{t_{2g},1} + 1 \big) 
\big[
( L^x_0 L^y_1 )^2 + 
( L^y_0 L^x_1 )^2 
\nonumber \\ 
&+&  L_0^x L_0^y L_1^x L_1^y + L_0^y L_0^x L_1^y L_1^x
\big]  
+ ( L_0^z )^2 + (L_1^z)^2
\Big],
\nonumber \\
\label{eq10}
\end{eqnarray}
where the components of ${\boldsymbol L}$ are defined in the $xyz$ coordinates in Fig.~\ref{fig1} and Fig.~\ref{fig3}. 
Again, this interaction is obtained by ignoring the direct hopping process between the $xy$ orbitals. The Hubbard
 interaction is assumed to be dominant compared to the spin-orbit coupling and the Hund's coupling such that
 the correction from the Hund's coupling in Eq.~\eqref{eq10} is not considered~\cite{PhysRevLett.111.197201}.  
The $(L^z)^2$ term in Eq.~\eqref{eq10}, after being added with the similar terms from the symmetry
equivalent bonds, will become constant, and thus can be neglected. 
Following our previous reasoning, 
the ${\mathcal O}$-${\mathcal O}$ interaction 
is identified as,
\begin{eqnarray}
&& J_{\text{KK}}\big( {\boldsymbol S}_{t_{2g},0} \cdot {\boldsymbol S}_{t_{2g},1} \big) \big[ 
( L^x_0 L^y_1 )^2 + 
( L^y_0 L^x_1 )^2 
\nonumber \\ 
&& \quad\quad\quad\quad\quad  
+  L_0^x L_0^y L_1^x L_1^y + L_0^y L_0^x L_1^y L_1^x
\big]  ,
\label{eq11}
\end{eqnarray}
and the ${\mathcal N}$-${\mathcal N}$ interaction is
identified as 
\begin{equation}
  J_{\text{KK}}  \big[ 
( L^x_0 L^y_1 )^2 + 
( L^y_0 L^x_1 )^2 
 +  L_0^x L_0^y L_1^x L_1^y + L_0^y L_0^x L_1^y L_1^x
\big] .
\label{eq12}
\end{equation}

\begin{widetext}
Ignoring the single-ion anisotropy $H_{\text{ani}}$, one can express the local $J$-states in terms
of the spin and orbital wavefunction in the local $xyz$ coordinate system.  
One then projects the above ${\mathcal O}$-${\mathcal O}$ and ${\mathcal N}$-${\mathcal N}$ interactions
onto the $J$-state basis. We then proceed and establish 
 the relationship between the operators ${\mathcal O}$ and the projected $J$-operators such as
\begin{eqnarray}
{\boldsymbol S} (L^x)^2 &=& (S^xL^xL^x, S^yL^xL^x, S^zL^xL^x) \xrightarrow[]{P_{J=1}} (\frac{6}{5}J^x, \frac{9}{10} J^y, \frac{9}{10} J^z ), 
\\
{\boldsymbol S} (L^y)^2 &=&  (S^xL^yL^y, S^yL^yL^y, S^zL^yL^y) \xrightarrow[]{P_{J=1}} (\frac{9}{10}J^x, \frac{6}{5} J^y, \frac{9}{10} J^z ), 
\\
{\boldsymbol S}L^x L^y &=& (S^xL^xL^y, S^yL^xL^y, S^zL^xL^y)  \xrightarrow[]{P_{J=1}} (\frac{3}{20} J^y-\frac{3i}{10} Q^{xz}, \frac{3}{20} J^x - \frac{3i}{10} Q^{yz} , -\frac{i}{2} -\frac{i\sqrt{3}}{10} Q^{3z^2}),
\\
{\boldsymbol S}L^y L^x &=&  (S^xL^yL^x, S^yL^yL^x, S^zL^yL^x)  \xrightarrow[]{P_{J=1}} 
(  \frac{3}{20} J^y +\frac{3i}{10} Q^{xz}, \frac{3}{20} J^x + \frac{3i}{10} Q^{yz} ,\frac{i}{2} +\frac{i\sqrt{3}}{10} Q^{3z^2} ) ,
\end{eqnarray}
where $P_{J=1}$ refers to the projection onto the ${J=1}$ manifold.

\end{widetext}

Using the above relations, we then obtain the quadrupole interaction from above ${\mathcal O}$-${\mathcal O}$ interaction
in Eq.~\eqref{eq11} that is given as
\begin{eqnarray}
&& -\frac{9J_{\text{KK}}}{200} \Big[ 
\frac{1}{3}Q_{0}^{3z^2}   Q_{1}^{3z^2}   
+ Q_0^{xz} Q_1^{xz}  + Q_0^{yz} Q_1^{yz}  
\Big].
\label{eq17}
\end{eqnarray}
Likewise, for the quadrupole interaction from above ${\mathcal N}$-${\mathcal N}$ interaction
in Eq.~\eqref{eq12}, we use the relations
\begin{eqnarray}
&& (L^x)^2  \xrightarrow[]{P_{J=1}} \frac{2}{3} +  \frac{1}{20} (- \frac{1}{\sqrt{3}}Q^{3z^2}  + Q^{x^2y^2} ),\\
&& (L^y)^2 \xrightarrow[]{P_{J=1}}  \frac{2}{3}+  \frac{1}{20} (- \frac{1}{\sqrt{3}} Q^{3z^2} - Q^{x^2y^2} ),\\
&& L^x L^y \xrightarrow[]{P_{J=1}} \frac{1}{10} Q^{xy}, \\
&& L^y L^x \xrightarrow[]{P_{J=1}} \frac{1}{10} Q^{xy},
\end{eqnarray}
and obtain
\begin{eqnarray}
\frac{J_{\text{KK}}}{200} \Big[ \frac{1}{3} Q_0^{3z^2} Q_1^{3z^2}  -  Q_0^{x^2y^2} Q_1^{x^2y^2} + 4 Q_0^{xy} Q_1^{xy}\Big]. 
\label{eq22}
\end{eqnarray}
The summation of Eq.~\eqref{eq17} and Eq.~\eqref{eq22} is the total quadrupole interaction for the 01 bond. 
Here we have expressed the quadrupole moments in the local $xyz$ coordinate for the simplicity of the expression. 
The relation between $Q$'s in the local coordinates and the $J^{x,y,z}$ is identical to the ones
for $Q$'s in the global $abc$ coordinates and the $J^{a,b,c}$.  
The relation of $Q$'s in the local $xyz$ coordinates and the global $abc$ coordinates are listed as
\begin{eqnarray}
&& Q^{3z^2} =  -\frac{\sqrt{3}}{3}Q^{a^2b^2} -\frac{2\sqrt{6}}{3} Q^{bc} , \\
&& Q^{x^2y^2} =   \frac{2\sqrt{3}}{3}Q^{ab} +\frac{2\sqrt{6}}{3} Q^{ac} , \\
&& Q^{xy} = - \frac{\sqrt{3} }{6} Q^{3c^2}  + \frac{Q^{a^2b^2}}{3}   - \frac{\sqrt{2}}{3} Q^{bc}   ,\\
&& {Q^{xz} =  \frac{\sqrt{3} }{6}  Q^{3c^2}+ \frac{Q^{a^2b^2}}{6}  -\frac{Q^{ab}}{\sqrt{3}}  + \frac{Q^{ac}}{\sqrt{6}}  - \frac{\sqrt{2}}{6}Q^{bc}} ,\\
&& {Q^{yz} =   -\frac{\sqrt{3}}{6} Q^{3c^2}- \frac{Q^{a^2b^2}}{6}   -\frac{ Q^{ab}}{\sqrt{3}} + \frac{Q^{ac}}{\sqrt{6}} +  \frac{\sqrt{2}}{6}Q^{bc} }
\end{eqnarray}
and the dipole moment $J^{\mu}$ is given as 
${J^a= \frac{1}{\sqrt{2}}(J^x+J^y)}$, ${J^b=\frac{1}{\sqrt{6}} (J^x-J^y-2J^z)}$ and ${J^c= \frac{1}{\sqrt{3}}(J^x - J^y + J^z)}$. 
This completes the derivation of the interaction between the quadrupole moment along the 
bond 01. For the other equivalent bonds, one can simply obtain the 
quadrupole interaction via the cubic permutation.


\emph{Discussion.}---Here we discuss the implication of the exchange interaction 
between the quadrupole moments for FeI$_2$. Clearly, 
the presence of the quadrupole interaction would enhance quantum fluctuations. 
This is because the quadrupole moment operators connect
all the spin states with more-or-less similar probability and allow the system
to tunnel quantum mechanically between different spin states more effectively~\cite{chen2021mott}. 
This aspect differs from the conventional dipole moment that only changes the spin 
quantum number by 0 or ${\pm 1}$ at one time. Thus, 
a system with the substantial quadrupole interaction is more likely to be more 
delocalized in its spin Hilbert space and favors more exotic quantum ground states,
and even for an ordered state, the usual spin-wave theory should be replaced by the flavor wave theory 
that takes into account of all possible channels connected by the dipole and quadrupole operators~\cite{Bai_2021,Li_2018,PhysRevB.100.045103,Liu_2020}.
   In Ref.~\onlinecite{Bai_2021}, however, the authors found a reasonable agreement with the inelastic neutron scattering
   spectroscopy based on a model with the single-ion anisotropy and the symmetry-allow
   pairwise spin interaction. It is likely that, the quadrupole interaction in FeI$_2$ might be weak 
   compared to the pairwise dipole interaction as the dipole interaction has more microscopic processes 
   than the quadrupole interaction according to the discussion. 
   Or, the quadrupole interaction does not impact the inelastic spectrum 
   significantly in FeI$_2$. The single-ion anisotropy, that is an onsite polarization 
   term of the $Q^{3c^2}$ moment, seems to be quite large in FeI$_2$~\cite{PhysRevLett.127.267201}. 
   This is expected from the presence of the several magnetization plateaux or steps for FeI$_2$
   in the $c$-direction magnetic field~\cite{PhysRevB.82.104402} as well as from 
   the recent time-domain terahertz spectroscopic measurement~\cite{PhysRevLett.127.267201}. 
   The presence of a large single-ion anisotropy 
   could suppress the effect of other quadrupole interactions. A more systematic analysis 
   may need to combine both the dipole and quadrupole interactions and at the same time 
   acquire a more quantitative input from the density functional theory for the exchange interactions. 
   The isostructural material NiI$_2$ has a fully-filled $t_{2g}$ manifold with quenched orbitals, and the 
   CoI$_2$ has a $t_{2g}^5$ configuration and falls into the well-known ${J=1/2}$ category~\cite{PhysRevB.87.014429}.

   Our work here provides a natural scheme to understand the origin of the multipolar interaction
   and can be particularly useful for the multi-orbital systems with distinct and separate orbital manifolds.  
   One relevant system would be the Co-based Kitaev materials that are under an active investigation recently. 
   In fact, for these Co-based honeycomb Kitaev magnets, the local moment for the Co$^{2+}$ ion 
   with a $3d^7$ electron 
   configuration on the octahedral environment can be understood as the combination of the spin-1 
   from the $e_g$ manifold and the spin-1/2 with an orbital moment ${L=1}$ from the $t_{2g}$ manifold. 
   The exchange interaction over there may be understood from the separation into the $e_g$ and the $t_{2g}$ manifolds. 
   It seems, a similar separation scheme was already been considered for the $3d^7$ cobaltates~\cite{PhysRevB.97.014407}.
   Since the local moment for Co$^{2+}$ is an effective ${J=1/2}$ moment, 
   the resulting interaction is always dipole interaction. This scheme 
   simply becomes a way to organize the superexchange processes, 
   but does not seem to provide much a simplification. 
   We thus expect a more useful application to occur in systems with larger-$J$ moments.



\emph{Acknowledgments.}---This work is supported by the Ministry of Science and Technology 
of China with Grants No. 2021YFA1400300, the National Science 
Foundation of China with Grant No. 92065203, and by the Research Grants Council of Hong Kong with C7012-21GF.

\bibliography{refs}

\end{document}